\documentclass[preprint,prl,superscriptaddress,showpacs]{revtex4}

\usepackage{graphicx}
\usepackage{amssymb}
\usepackage{amsfonts}

\DeclareGraphicsExtensions{.eps}

\begin{document}

\title{Twin polaritons in semiconductor microcavities}

\date{\today}

\author{J.Ph. Karr}
\affiliation{Laboratoire Kastler Brossel, Universit\'{e} Paris 6, Ecole
Normale Sup\'{e}rieure et CNRS,\\
UPMC Case 74, 4 place Jussieu, 75252 Paris Cedex 05, France}

\author{A. Baas}
\affiliation{Laboratoire Kastler Brossel, Universit\'{e} Paris 6, Ecole
Normale Sup\'{e}rieure et CNRS,\\
UPMC Case 74, 4 place Jussieu, 75252 Paris Cedex 05, France}

\author{E. Giacobino}
\affiliation{Laboratoire Kastler Brossel, Universit\'{e} Paris 6, Ecole
Normale Sup\'{e}rieure et CNRS,\\
UPMC Case 74, 4 place Jussieu, 75252 Paris Cedex 05, France}

\begin{abstract}
The quantum correlations between the beams generated by polariton pair scattering in a semiconductor microcavity above the
parametric oscillation threshold are computed analytically. The influence of various parameters like the cavity-exciton
detuning, the intensity mismatch between the signal and idler beams and the amount of spurious noise is analyzed. We show
that very strong quantum correlations between the signal and idler polaritons can be achieved. The quantum effects on the
outgoing light fields are strongly reduced due to the large mismatch in the coupling of the signal and idler polaritons to
the external photons.
\end{abstract}


\maketitle

\section{introduction}

\begin{figure}[p]
\centerline{\includegraphics[clip=,width=8cm]{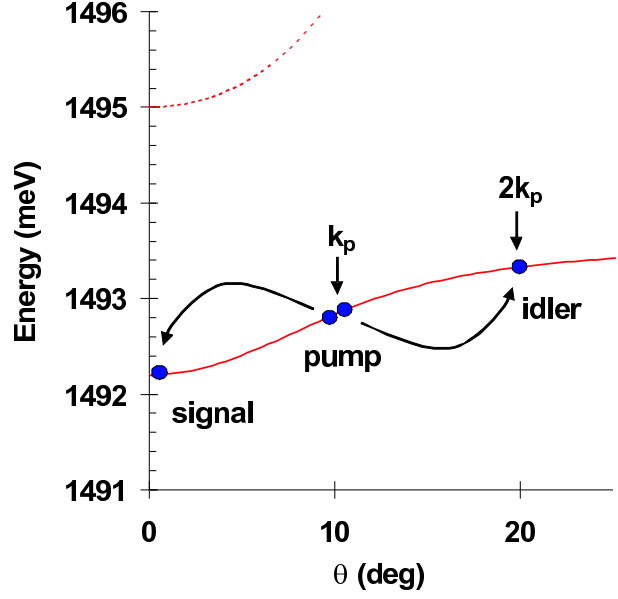}} \caption{Energy dispersion of the two polariton branches for a
microcavity sample having a Rabi splitting of 2.8 meV, at zero cavity-exciton detuning. The arrows show the parametric
conversion of the pump polaritons ($\simeq 10^{\circ}$) into signal ($0^{\circ}$) and idler ($\simeq 20^{\circ}$)
polaritons.} \label{magic}
\end{figure}

Exciton polaritons are the normal modes of the strong light-matter coupling in semiconductor microcavities
\cite{weisbuch}. These half-exciton, half-photon particles present large optical nonlinearities coming from the Coulomb
interactions between the exciton components. Under resonant pumping, this leads to a parametric process where a pair of
pump polaritons scatter into nondegenerate signal and idler modes while conserving energy and momentum. The scattering is
particularly strong in microcavities because the unusual shape of the polariton dispersion makes it possible for the pump,
signal and idler to be on resonance at the same time (see Fig.~\ref{magic}). Moreover, the relationship between the
in-plane momentum of each polariton mode and the direction of the external photon to which it couples \cite{houdre}
enables to investigate the parametric scattering using measurements at different angles to access the various modes.

The first demonstration of parametric processes in semiconductor microcavities was performed by Savvidis \textit{et al.}
\cite{savvidis} using ultrafast pump-probe measurements. He observed parametric amplification, where the scattering is
stimulated by excitation of the signal mode with a weak probe field. Parametric oscillation, where there is no probe and a
coherent population in the signal and idler modes appears spontaneously, has since been observed by Stevenson \textit{et
al.} \cite{stevenson} and Baumberg \textit{et al.} \cite{baumberg} in cw experiments. The lower polariton was pumped
resonantly at the "magic" angle of about 16$^{\circ}$. Above a threshold pump intensity, strong signal and idler beams
were observed at about 0$^{\circ}$ and 35$^{\circ}$, without any probe stimulation. The coherence of these beams was
demonstrated by a significant spectral narrowing.

The large optical nonlinearity of cavity polaritons makes them very attractive for quantum optics. Noise reduction on the
reflected light field has been predicted \cite{messin} and achieved experimentally \cite{squeezing} for a resonant pumping
of the lower polariton at 0$^{\circ}$. The parametric fluorescence was recently shown to produce strongly correlated pairs
of signal and idler polaritons, yielding a two-mode squeezed state \cite{quattropani}. The parametric oscillation regime
is also very interesting in this respect \cite{karr}. It is well known that optical parametric oscillators (OPO) can be
used to generate twin beams, the fluctuations of which are correlated at the quantum level. A noise reduction of 86\% was
obtained by substracting the intensities of the signal and idler beams produced by a LiNbO$_{3}$ OPO \cite{mertz}.

The purpose of this paper is to investigate the possibility of generating twin beams using a semiconductor microcavity
above the parametric oscillation threshold. The classical model developed by Whittaker \cite{whittaker} is no longer
sufficient to study the quantum noise properties of the system. Thus we adapt the quantum model by Ciuti \textit{et al.},
previously used in the context of parametric amplification \cite{ciuti2} and parametric fluorescence
\cite{ciuti3,quattropani}, to the parametric oscillator configuration. Furthermore we compute the field fluctuations using
the input-output method \cite{collett,reynaud}. We also include the excess noise associated with the excitonic relaxation,
which had not been done in Refs. \cite{ciuti3,quattropani}.

\section{Model}

\subsection{Hamiltonian}

Following Ciuti \textit{et al.} \cite{ciuti2}\cite{ciuti3} we write the effective Hamiltonian for the coupled
exciton-photon system. The spin degree of freedom is neglected.

\begin{equation}
H=H_{0}+H_{exc-exc}+H_{sat} \label{h}
\end{equation}

The first term is the linear Hamiltonian for excitons and cavity photons

\begin{eqnarray}
H_{0} &=& \sum_{\mathbf{k}} E_{exc}(k) b_{\mathbf{k}}^{\dagger} b_{\mathbf{k}} + \sum_{\mathbf{k}} E_{cav}(k)
a_{\mathbf{k}}^{\dagger} a_{\mathbf{k}} \nonumber \\&& + \sum_{\mathbf{k}} \hbar \Omega_{R} \left(
a_{\mathbf{k}}^{\dagger} b_{\mathbf{k}} + b_{\mathbf{k}}^{\dagger} a_{\mathbf{k}} \right)
\end{eqnarray}

with $b_{\mathbf{k}}^{\dagger}$ and $a_{\mathbf{k}}^{\dagger}$ the creation operators respectively for excitons and
photons of in-plane wave vector $\mathbf{k}$, which satisfy boson commutation rules. $E_{exc}(k)$ and $E_{cav}(k)$ are the
energy dispersions for exciton and cavity mode. The last term represents the linear coupling between exciton and cavity
photon which causes the vacuum Rabi splitting $2\hbar \Omega_{R}$. The fermionic nature of electrons and holes causes a
deviation of the excitons from bosonic behavior, which is accounted for through an effective exciton-exciton interaction
and exciton saturation. The exciton-exciton interaction term writes

\begin{equation}
H_{exc-exc}=\frac{1}{2}\sum\limits_{\mathbf{k},\mathbf{k'},\mathbf{q}}V_{q} b_{\mathbf{k}+\mathbf{q}}^{\dagger}
b_{\mathbf{k'}-\mathbf{q}}^{\dagger} b_{\mathbf{k}}b_{\mathbf{k'}}
\end{equation}

where $V_{q}\simeq V_{0}=\frac{6e^{2}a_{exc}}{\epsilon _{0}A}$ for $qa_{exc}\ll 1$, $a_{exc}$ being the two-dimensional
exciton Bohr radius, $\epsilon _{0}$ the dielectric constant of the quantum well and $A$ the macroscopic quantization
area. The saturation term in the light-exciton coupling is

\begin{equation}
H_{sat}=-\sum\limits_{\mathbf{k},\mathbf{k'},\mathbf{q}} V_{sat} \left( a_{\mathbf{k}+\mathbf{q}}^{\dagger
}b_{\mathbf{k'}-\mathbf{q}}^{\dagger }b_{\mathbf{k}}b_{\mathbf{k'}}+a_{\mathbf{k}+\mathbf{q}}
b_{\mathbf{k'}-\mathbf{q}}b_{\mathbf{k}}^{\dagger }b_{\mathbf{k'}}^{\dagger }\right)
\end{equation}

where $V_{sat}=\frac{\hbar \Omega_{R} }{n_{sat}A}$ with $n_{sat}=7/\left( 16\pi a_{exc}^{2}\right)$ being the exciton
saturation density. We consider resonant or quasi-resonant excitation of the lower polariton branch by a
quasi-monochromatic laser field of frequency $\omega_{L}=E_{L}/\hbar$ and wave vector $\mathbf{k_{L}}$. If the pump
intensity is not too high the resonances (i.e. the polariton states) are not modified. Then it is much more convenient to
work directly in the polariton basis. It is possible to neglect nonlinear contributions related to the upper branch and
consider only the lower polariton states. The polariton operators are obtained by a unitary transformation of the exciton
and photon operators:

\begin{equation}
\left( \begin{array}{c}
p_{\mathbf{k}}\\
q_{\mathbf{k}}
\end{array} \right)=
\left( \begin{array}{cc}
-C_{k} & X_{k} \\
 X_{k} & C_{k}
\end{array} \right)
\left( \begin{array}{c}
a_{\mathbf{k}} \\
b_{\mathbf{k}}
\end{array} \right)
\label{defpolaritons}
\end{equation}

where $X_{k}$ et $C_{k}$ are positive real numbers called the Hopfield coefficients, given by

\begin{eqnarray}
X_{k}^{2} &=&\frac{\delta _{k}+\sqrt{\delta _{k}^{2}+\Omega_{R} ^{2}}}{2\sqrt{%
\delta _{k}^{2}+\Omega_{R} ^{2}}} \\
C_{k}^{2} &=&\frac{\Omega_{R} ^{2}}{2\sqrt{\delta _{k}^{2}+\Omega_{R} ^{2}}\left( \delta _{k}+\sqrt{\delta
_{k}^{2}+\Omega_{R} ^{2}}\right) } \label{hopfield}
\end{eqnarray}

$X_{k}^{2}$ and $C_{k}^{2}$ can be interpreted respectively as the exciton and photon fraction of the lower polariton
$p_{\mathbf{k}}$. In terms of the lower polariton operators, the Hamiltonian (\ref{h}) reads

\begin{equation}
H=H_{P}+H_{PP}^{eff} \label{heff}
\end{equation}

$H_{P}$ is the free evolution term for the lower polariton:

\begin{equation}
H_{P}=\sum\limits_{\mathbf{k}} E_{P} \left( k \right) p_{\mathbf{k}}^{\dagger }p_{\mathbf{k}}
\end{equation}

and $H_{PP}^{eff}$ is an effective polariton-polariton interaction:

\begin{equation}
H_{PP}^{eff}=\frac{1}{2}\sum\limits_{\mathbf{k},\mathbf{k^{\prime}},\mathbf{q}}
V_{\mathbf{k},\mathbf{k^{\prime}},\mathbf{q}}^{PP} p_{\mathbf{k+q}}^{\dagger }p_{\mathbf{k^{\prime }-q}}^{\dagger
}p_{\mathbf{k}}p_{\mathbf{k^{\prime}}}
\end{equation}

where

\begin{eqnarray}
V_{\mathbf{k},\mathbf{k^{\prime}},\mathbf{q}}^{PP} &=& \left\{ V_{0} X_{|\mathbf{k+q}|} X_{k^{\prime }}+ 2 V_{sat} \right.
\\&& \left. \times \left( C_{|\mathbf{k+q}|} X_{k^{\prime}} + C_{k^{\prime}} X_{|\mathbf{k+q}|} \right) \right\}
X_{|\mathbf{k^{\prime}-q}|} X_{k} \nonumber
\end{eqnarray}

In the following we neglect the contribution of the saturation term, so that
$V_{\mathbf{k},\mathbf{k^{\prime}},\mathbf{q}}^{PP} \simeq V_{0} X_{|\mathbf{k+q}|} X_{k^{\prime}}
X_{|\mathbf{k^{\prime}-q}|} X_{k}$. We also neglect multiple diffusions i.e. interaction between modes other than the pump
mode. This approximation is valid only slightly above the parametric oscillation threshold \footnote{Multiple diffusions
were demonstrated in Refs. \cite{savvidis2,tarta}}. It comes to considering only the terms where the pump polariton
operator $p_{\mathbf{k_{L}}}$ appears at least twice:

\begin{eqnarray}
H_{PP}^{eff} &=& \frac{1}{2} \; V_{\mathbf{k_{L}},\mathbf{k_{L}},\mathbf{0}} p_{\mathbf{k_{L}}}^{\dagger}
p_{\mathbf{k_{L}}}^{\dagger} p_{\mathbf{k_{L}}} p_{\mathbf{k_{L}}} \nonumber \\&& + \sum\limits_{\mathbf{k}\neq
\mathbf{k_{L}}} V_{\mathbf{k_{L}},\mathbf{k_{L}},\mathbf{k_{L}-k}} \left(p_{2\mathbf{k_{L}}-\mathbf{k}}^{\dagger }
p_{\mathbf{k}}^{\dagger } p_{\mathbf{k_{L}}} p_{\mathbf{k_{L}}} + h.c. \right) \nonumber \\&& + \; 2 \;
\sum\limits_{\mathbf{k}\neq \mathbf{k_{L}}} V_{\mathbf{k},\mathbf{k_{L}},\mathbf{0}}^{PP} p_{\mathbf{k_{L}}}^{\dagger }
p_{\mathbf{k}}^{\dagger} p_{\mathbf{k_{L}}} p_{\mathbf{k}} \label{nonlineaire}
\end{eqnarray}

The first term is a Kerr-like term for the polaritons in the pump mode. The second term is a "fission" process, where two
polaritons of wave vector $\mathbf{k_{L}}$ are converted into a "signal" polariton of wave vector $\mathbf{k}$ and an
"idler" polariton of wave vector $2\mathbf{k_{L}}-\mathbf{k}$. The last term corresponds to the interaction of the pump
mode $\mathbf{k_{L}}$ with all the other $\mathbf{k}$ states, which results in a blueshift proportional to $\left|
p_{\mathbf{k_{L}}}\right| ^{2}$.

\subsection{Energy conservation}

The energy conservation for the fission process $\{\mathbf{k_{L}},\mathbf{k_{L}}\} \rightarrow
\{\mathbf{k},\mathbf{2k_{L}-k}\}$ reads

\begin{equation}
\widetilde{E}_{P}\left( \mathbf{k}\right) +\widetilde{E}_{P}\left( 2\mathbf{k_{L}}-\mathbf{k}\right) =2
\widetilde{E}_{P}\left( \mathbf{k_{L}}\right) \label{conservation energie}
\end{equation}

where $\widetilde{E}_{P}\left( \mathbf{q}\right)$ is the energy of the polariton of wave vector $\mathbf{q}$, renormalized
by the interaction with the pump polaritons

\begin{equation}
\widetilde{E}_{P}\left( \mathbf{q}\right) = E_{P}\left( q\right) +2V_{\mathbf{q},\mathbf{k_{L}},\mathbf{0}}\left|
\left\langle p_{\mathbf{k_{L}}} \right\rangle \right| ^{2} \label{renorm}
\end{equation}

\begin{figure}[p]
\centerline{\includegraphics[clip=,width=8cm]{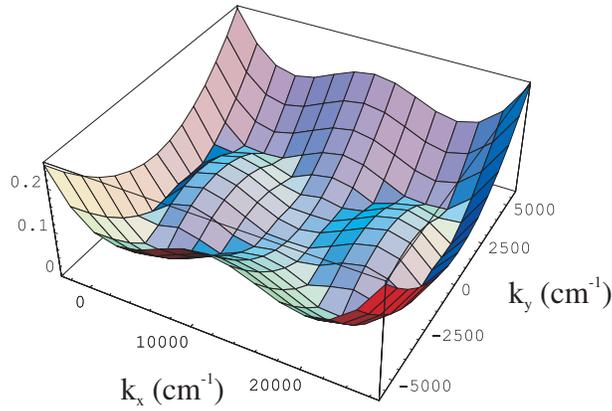}} \caption{Plot of the quantity $\left| E_{P}\left(
\mathbf{k}\right) +E_{P}\left( 2\mathbf{k_{L}}-\mathbf{k}\right) - 2 E_{P}\left( \mathbf{k_{L}}\right) \right|$ (in meV)
as a function of $k_{x}$ and $k_{y}$ (in cm$^{-1}$), for the parameters of Fig.~\ref{magic}.} \label{consenergie}
\end{figure}

Note that the factor of 2 disappears for $\mathbf{q}=\mathbf{k_{L}}$. Equation (\ref{conservation energie}) always has a
trivial solution $\mathbf{k}=\mathbf{k_{L}}$. Non-trivial solutions exist provided the wave vector $\mathbf{k_{L}}$ is
above a critical value, or equivalently if the angle of incidence is above the so-called "magic angle" $\theta_{c}$
\cite{savvidis}. From now on we suppose that the microcavity is excited resonantly with an angle $\theta_{c}$.
Fig.~\ref{consenergie} is a plot of the quantity $\left| E_{P}\left( \mathbf{k}\right) +E_{P}\left(
2\mathbf{k_{L}}-\mathbf{k}\right) - 2 E_{P}\left( \mathbf{k_{L}}\right) \right|$ as a function of $\mathbf{k}=\left\{
k_{x},k_{y} \right\}$, $\mathbf{k_{L}}$ being parallel to the $x$ axis.

This shows that energy conservation can be satisfied for a wide range of wave vectors
$\{\mathbf{k},2\mathbf{k_{L}}-\mathbf{k}\}$. In recent experiments, parametric oscillation was observed in the normal
direction $\mathbf{k}=0$ \cite{stevenson,houdre2}. In this paper we consider only the parametric process $\left\{
\mathbf{k_{L}},\mathbf{k_{L}}\right\} \longrightarrow \{\mathbf{0},2\mathbf{k_{L}}\}$ assuming that the other ones remain
below threshold. Then we can neglect the effect of modes other than $\mathbf{0},\mathbf{k_{L}},2\mathbf{k_{L}}$. The
evolution of these three modes is given by a closed set of equations.

\section{Heisenberg-Langevin equations}

In order to study the quantum fluctuations we have to write the Heisenberg-Langevin equations including the relaxation and
fluctuation terms. The relaxation of the cavity mode comes from the interaction with the external electromagnetic field
through the Hamiltonian

\begin{equation}
H_{I} = i \hbar \int \frac{d\omega}{2 \pi} \kappa \left( a_{\mathbf{k}}^{\dagger} A_{\omega} - A_{\omega}^{\dagger}
a_{\mathbf{k}} \right)
\end{equation}

The coupling constant is given by $\kappa=\sqrt{2 \gamma_{ak}}$ where $\gamma_{ak}$ is the cavity linewidth (HWHM). This
leads to the following evolution equation for the cavity field:

\begin{equation}
\frac{da_{\mathbf{k}}}{dt}(t)=- \gamma_{ak} a_{\mathbf{k}}(t) + \sqrt {2\gamma_{ak}} A_{\mathbf{k}}^{in}(t)
\label{evolutiona}
\end{equation}

where $A_{\mathbf{k}}^{in}(t)$ is the incoming external field. In this equation the normalization are not the same for the
cavity field as for the external field: $n_{a_{\mathbf{k}}}(t)=\left\langle a_{\mathbf{k}}^{\dagger}(t) a_{\mathbf{k}}(t)
\right\rangle$ is the mean number of cavity photons, while $I_{\mathbf{k}}^{in}=\left\langle A_{\mathbf{k}}^{in\dagger}(t)
A_{\mathbf{k}}^{in}(t) \right\rangle$ is the mean number of incident photons per second.

Exciton relaxation is a much more complex problem. The density is assumed to be low enough to neglect the relaxation due
to exciton-exciton interaction \cite{ciuti98}. At low density and low enough temperature the main relaxation mechanism is
the interaction with acoustic phonons. A given exciton mode $b_{\mathbf{k}}$ is coupled to all the other exciton modes
$b_{\mathbf{k'}}$ and to all the phonon modes fulfilling the condition of energy and wavevector conservation
\cite{piermarocchi}. Relaxation in microcavities in the strong coupling regime has been studied in detail
\cite{tassone96,tassone99}. However, the derivation of the corresponding fluctuation terms requires additional hypotheses,
under which one can replace the exciton-phonon coupling Hamiltonian by a linear coupling to a single reservoir
\cite{karr}. Then, in the same way as for the photon field, the fluctuation-dissipation part in the Langevin equation for
the excitons writes

\begin{equation}
\frac{db_{\mathbf{k}}(t)}{dt}=-\gamma _{bk}b_{\mathbf{k}}(t)+\sqrt{2\gamma _{bk}} B_{\mathbf{k}}^{in}(t)
\end{equation}

where $\gamma _{bk}$ is the exciton linewidth (HWHM) and $B_{\mathbf{k}}^{in}(t)$ the input excitonic field, which is a
linear combination of the reservoir modes.

Using these results we can write the Heisenberg-Langevin equations for the cavity and exciton modes of wave vectors
$\mathbf{0},\mathbf{k_{L}},2\mathbf{k_{L}}$ and then for the three corresponding lower polariton modes. We define the
slowly varying operators

\begin{eqnarray}
\widetilde{p}_{\mathbf{k_{L}}}(t) &=& p_{\mathbf{k_{L}}}(t) e^{iE_{L}t/\hbar} \nonumber \\
\widetilde{p}_{\mathbf{0}}(t) &=& p_{\mathbf{0}}(t) e^{iE_{p}(0)t/\hbar} \nonumber \\
\widetilde{p}_{2\mathbf{k_{L}}}(t) &=& p_{2\mathbf{k_{L}}}(t) e^{iE_{p}(2k_{L})t/\hbar} \label{tournant2}
\end{eqnarray}

which obey the following equations:

\begin{eqnarray}
\frac{d\widetilde{p}_{\mathbf{0}}}{dt} &=& -\frac{i}{\hbar }\left( 2 V_{\mathbf{0},\mathbf{k_{L}},\mathbf{0}}
\widetilde{p}_{\mathbf{k_{L}}}^{\dagger}
\widetilde{p}_{\mathbf{k_{L}}} - i\gamma _{0}\right) \widetilde{p}_{\mathbf{0}} \label{eqsignal} \\
&& - \frac{i }{\hbar } V_{\mathbf{k_{L}},\mathbf{k_{L}},\mathbf{k_{L}}} \widetilde{p}_{2\mathbf{k_{L}}}^{\dagger}
\widetilde{p}_{\mathbf{k_{L}}}^{2}
e^{i\Delta E t/\hbar} + P_{\mathbf{0}}^{in} \nonumber \\
\frac{d\widetilde{p}_{2\mathbf{k_{L}}}}{dt} &=& -\frac{i}{\hbar }\left( 2 V_{2\mathbf{k_{L}},\mathbf{k_{L}},\mathbf{0}}
 \widetilde{p}_{\mathbf{k_{L}}}^{\dagger} \widetilde{p}_{\mathbf{k_{L}}}-i\gamma _{2 k_{L}}\right)
 \widetilde{p}_{2\mathbf{k_{L}}} \label{eqcomplementaire}\\
&&-\frac{i}{\hbar} V_{\mathbf{k_{L}},\mathbf{k_{L}},\mathbf{k_{L}}} \widetilde{p}_{\mathbf{0}}^{\dagger}
\widetilde{p}_{\mathbf{k_{L}}}^{2} e^{i\Delta E
t/\hbar} + P_{2\mathbf{k_{L}}}^{in} \nonumber \\
\frac{d\widetilde{p}_{\mathbf{k_{L}}}}{dt} &=& -\frac{i}{\hbar}\left( \Delta_{L} +
V_{\mathbf{k_{L}},\mathbf{k_{L}},\mathbf{0}} \widetilde{p}_{\mathbf{k_{L}}}^{\dagger} \widetilde{p}_{\mathbf{k_{L}}}
-i\gamma _{k_{L}}\right) \widetilde{p}_{\mathbf{k_{L}}} \label{eqpompe} \\
&&-\frac{2i}{\hbar }V_{\mathbf{k_{L}},\mathbf{k_{L}},\mathbf{k_{L}}}\widetilde{p}_{\mathbf{k_{L}}}^{\dagger
}\widetilde{p}_{\mathbf{0}}\widetilde{p}_{2\mathbf{k_{L}}} e^{-i\Delta E t/\hbar} + P_{\mathbf{k_{L}}}^{in} \nonumber
\end{eqnarray}

where for any given wave vector $\mathbf{q}$, $P_{\mathbf{q}}^{in} = - C_{q} \sqrt{2\gamma_{aq}}A_{\mathbf{q}}^{in}
+X_{q}\sqrt{2\gamma _{bq}}B_{\mathbf{q}}^{in}$ is the polariton input field (which is a linear combination of the cavity
and exciton input fields ; only the driving laser field $A_{\mathbf{k_{L}}}^{in}$ has a nonzero mean value), $\gamma
_{q}=C_{q}^{2} \gamma_{aq}+X_{q}^{2} \gamma_{bq}$ is the polariton linewitdth ; $\Delta_{L}=E_{p}(k_{L})-E_{L}$ is the
laser detuning ; $\Delta E = E_{p}(2k_{L})+E_{p}(0)-2E_{L}$ is the energy mismatch.

These equations extend the model developed by Ciuti \textit{et al.} \cite{ciuti3} above threshold. The full treatment of
the field fluctuations is included as well as the equation of motion of the pumped mode accounting for the pump depletion.
They are valid only slightly above threshold, because far above threshold when we can no longer neglect multiple
scattering \cite{savvidis2,tarta}.

This set of equations is similar to the evolution equations of a non-degenerate optical parametric oscillator (OPO)
\cite{fabre}. The non-linearity is of type $\chi^{(3)}$ while in most OPOs it is of type $\chi^{(2)}$. OPOs based on
four-wave mixing have already been demonstrated \cite{michel}. However, let us stress that here the parametric process
involves the excitations of a semiconductor material (i.e. polaritons) instead of photons. In the following we evaluate
the potential applications of this new type of OPO in quantum optics. The hybrid nature of polaritons makes the treatment
of quantum fluctuations more complicated, since we have to consider additional sources of noise (i.e. the luminescence of
excitons).

\section{Mean fields above threshold}

To start with we have to compute the stationary state of the system. This comes to the calculation done by Whittaker in
Ref. \cite{whittaker}. We neglect the renormalization effects due to the interaction with the pump mode, which allows to
get analytical expressions. Moreover, we suppose that the angle of incidence is adjusted in order to satisfy the resonance
condition $\Delta E = 0$ and that the pump laser is perfectly resonant ($\Delta_{L}$=0). Equations
(\ref{eqsignal})-(\ref{eqpompe}) now write

\begin{eqnarray}
\frac{d\widetilde{p}_{\mathbf{0}}}{dt} &=& -\gamma _{0}\widetilde{p}_{\mathbf{0}} -
iE_{int}\widetilde{p}_{2\mathbf{k_{L}}}^{\dagger
}\widetilde{p}_{\mathbf{k_{L}}}^{2} + P_{\mathbf{0}}^{in}  \label{evolsignal} \\
\frac{d\widetilde{p}_{2\mathbf{k_{L}}}}{dt} &=& -\gamma _{2\mathbf{k_{L}}}\widetilde{p}_{2\mathbf{k_{L}}} -
iE_{int}\widetilde{p}_{\mathbf{0}}^{\dagger
}\widetilde{p}_{\mathbf{k_{L}}}^{2} + P_{2\mathbf{k_{L}}}^{in} \label{evolcompl} \\
\frac{d\widetilde{p}_{\mathbf{k_{L}}}}{dt} &=& -\gamma _{\mathbf{k_{L}}}\widetilde{p}_{\mathbf{k_{L}}} -
2iE_{int}\widetilde{p}_{\mathbf{k_{L}}}^{\dagger }\widetilde{p}_{\mathbf{0}} \widetilde{p}_{2\mathbf{k_{L}}}+
P_{\mathbf{k_{L}}}^{in} \label{evolpompe}
\end{eqnarray}

where $E_{int}= V_{\mathbf{k_{L}},\mathbf{k_{L}},\mathbf{k_{L}}} / \hbar$. Let us recall that among the polariton input
fields, only the photon part of $P_{\mathbf{k_{L}}}^{in}$ corresponding to the pump laser field has a nonzero mean value.
The excitonic input fields $B_{\mathbf{q}}^{in}$ correspond to the luminescence of the exciton modes and are incoherent
fields with zero mean value. The stationary state is given by

\begin{eqnarray}
-\gamma _{k_{L}}\overline{p}_{\mathbf{k_{L}}}-2iE_{int}\overline{p}_{\mathbf{k_{L}}}^{\ast }
\overline{p}_{\mathbf{0}}\overline{p}_{2\mathbf{k_{L}}} &=&C_{k_{L}}\sqrt{2\gamma _{a}}
\overline{A}_{\mathbf{k_{L}}}^{in}  \label{stat pompe} \\
-\gamma _{0}\overline{p}_{\mathbf{0}}-iE_{int}\overline{p}_{2\mathbf{k_{L}}}^{\ast }\overline{p
}_{\mathbf{k_{L}}}^{2} &=&0  \label{stat signal} \\
-\gamma _{2k_{L}}\overline{p}_{2\mathbf{k_{L}}}^{\ast }+iE_{int}\overline{p}_{\mathbf{0}}
\overline{p}_{\mathbf{k_{L}}}^{\ast 2} &=&0  \label{stat compl}
\end{eqnarray}

For a non trivial solution to exist, the determinant of the last two equations must be zero:

\begin{equation}
E_{int}^{2}\left| \overline{p}_{\mathbf{k_{L}}}\right| ^{4}-\gamma _{0}\gamma _{2k_{L}}=0
\end{equation}

which gives the pump polariton population threshold

\begin{equation}
\left| \overline{p}_{\mathbf{k_{L}}}\right| ^{2}=\frac{\sqrt{\gamma _{0}\gamma _{2k_{L}}}}{E_{int}} \label{intensite
seuil}
\end{equation}

and the pump intensity threshold

\begin{equation}
I_{\mathbf{k_{L}},S}^{in}=\left| \overline{A}_{\mathbf{k_{L}},S}^{in}\right| ^{2}=\frac{\gamma _{k_{L}}^{2}\left( \gamma
_{0}\gamma _{2k_{L}}\right) ^{1/2}}{2\gamma _{a}C_{k_{L}}^{2}E_{int}}
\end{equation}

The signal and idler polariton populations are easily derived:

\begin{eqnarray}
\left| \overline{p}_{\mathbf{0}}\right| ^{2} &=&\frac{\gamma _{k_{L}}}{2E_{int}}\sqrt{%
\frac{\gamma _{2k_{L}}}{\gamma _{0}}}\left( \sigma -1\right)
\label{intensite signal} \\
\left| \overline{p}_{2\mathbf{k_{L}}}\right| ^{2} &=&\frac{\gamma _{k_{L}}}{2E_{int}}%
\sqrt{\frac{\gamma _{0}}{\gamma _{2k_{L}}}}\left( \sigma -1\right) \label{intensite compl}
\end{eqnarray}

where $\sigma =\sqrt{I_{\mathbf{k_{L}}}^{in}/I_{\mathbf{k_{L}},S}^{in}}$ is the pump parameter. We finally get the
intensities of the signal and idler output light fields

\begin{eqnarray}
I_{\mathbf{0}}^{out} &=&2\gamma _{a}C_{0}^{2}\left| \overline{p}_{\mathbf{0}}\right| ^{2}= \frac{\gamma _{a}\gamma
_{k_{L}}C_{0}^{2}}{E_{int}}\sqrt{\frac{\gamma
_{2k_{L}}}{\gamma _{0}}}\left( \sigma -1\right) \\
I_{2\mathbf{k_{L}}}^{out} &=&2\gamma _{a}C_{2k_{L}}^{2}\left| \overline{p} _{2\mathbf{k_{L}}}\right| ^{2}=\frac{\gamma
_{a}\gamma _{k_{L}}C_{2k_{L}}^{2}}{E_{int} }\sqrt{\frac{\gamma _{0}}{\gamma _{2k_{L}}}}\left( \sigma -1\right) \nonumber
\end{eqnarray}

Above  threshold, all the polaritons created by the pump are transferred to the signal and idler modes, so that the number
of pump polaritons is clamped to a fixed value. This phenomenon called pump depletion is well-known in triply resonant
OPOs. The signal and idler intensities grow like $\sqrt{I_{\mathbf{k_{L}}}^{in}}$. These results are in agreement with
those of Ref. \cite{whittaker}.

Finally we study the ratio of the signal and idler output intensities, which is an important parameter in view of the
analysis of the correlations between these two beams. It is given by the simple equation

\begin{equation}
\frac{I_{\mathbf{0}}^{out}}{I_{2\mathbf{k_{L}}}^{out}}=\frac{\gamma _{2k_{L}}C_{0}^{2}}{\gamma _{0}C_{2k_{L}}^{2}}
\label{rapportint}
\end{equation}

\begin{figure}[p]
\centerline{\includegraphics[clip=,width=8cm]{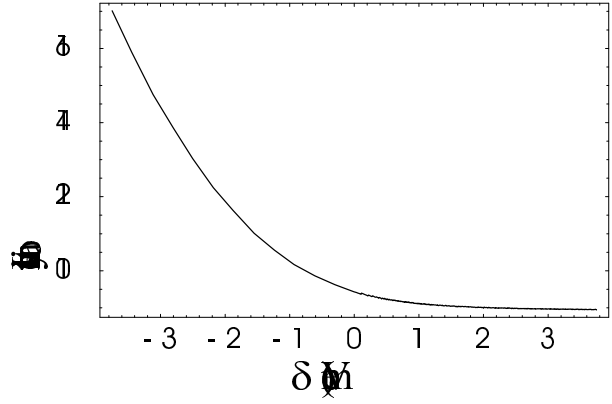}} \caption{Ratio of the photonic fractions of the signal and idler
polaritons as a function of the cavity-exciton detuning $\delta$. The Rabi splitting is 2.8 meV.} \label{ratio}
\end{figure}

We consider a typical III-V microcavity sample containing one quantum well, with a Rabi splitting $2\hbar\Omega_{R}$= 2.8
meV. At zero cavity-exciton detuning, one finds $k_{L}$=1.15 10$^{4}$ cm$^{-1}$. The photon fractions of the signal and
idler modes are respectively $C_{0}^{2}$=0.5 and $C_{2k_{L}}^{2} \simeq$ 0.053. Assuming that they have equal linewidths
the signal beam power should be about ten times that of the idler beam. It is possible to reduce this ratio a bit by
increasing the cavity-exciton detuning, as can be seen in Fig~\ref{ratio}. However, the oscillation threshold goes up. In
the following, all the results will be given at zero detuning.

\section{Fluctuations}

\subsection{Linearized evolution equations}

For any operator $O(t)$ we define a fluctuation operator $\delta O(t)=O(t)-\left\langle O(t) \right\rangle$. In order to
compute the fluctuations, we use the "semiclassical" linear input-output method, which consists in studying the
transformation of the incident fluctuations by the system \cite{reynaud}. It has been shown to be equivalent to a full
quantum treatment. We linearize equations (\ref{evolsignal})-(\ref{evolpompe}) in the vicinity of the working point
$p_{0}$ computed in the previous section. We obtain the following set of equations:

\begin{eqnarray}
\frac{d\delta p_{\mathbf{k_{L}}}}{dt} &=& -\gamma _{\mathbf{k_{L}}}\delta p_{\mathbf{k_{L}}} - 2iE_{int}\left(
\overline{p}_{\mathbf{0}}\overline{p}_{2\mathbf{k_{L}}}\delta p_{\mathbf{k_{L}}}^{\dagger } \right. \\&& \left. +
\overline{p}_{\mathbf{k_{L}}}^{\ast }\overline{p}_{2\mathbf{k_{L}}}\delta p_{\mathbf{0}} +
\overline{p}_{\mathbf{k_{L}}}^{\ast }\overline{p}_{\mathbf{0}}\delta p_{2\mathbf{k_{L}}}\right)
+ \delta P_{\mathbf{k_{L}}}^{in} \nonumber \\
\frac{d\delta p_{\mathbf{0}}}{dt} &=& -\gamma _{0}\delta p_{\mathbf{0}} - iE_{int}\left(
2\overline{p}_{2\mathbf{k_{L}}}^{\ast }\overline{p}_{\mathbf{k_{L}}}\delta p_{\mathbf{k_{L}}} \right. \\&& \left.
\overline{p }_{\mathbf{k_{L}}}^{2}\delta p_{2\mathbf{k_{L}}}^{\dagger }\right)
+ \delta P_{\mathbf{0}}^{in} \nonumber \\
\frac{d\delta p_{2\mathbf{k_{L}}}}{dt} &=& -\gamma _{2k_{L}}\delta p_{2\mathbf{k_{L}}}-iE_{int}\left(
2\overline{p}_{\mathbf{0}}^{\ast }\overline{p} _{\mathbf{k_{L}}}\delta p_{\mathbf{k_{L}}} \right. \\&& \left.
\overline{p}_{\mathbf{k_{L}}}^{2}\delta p_{\mathbf{0}}^{\dagger}\right) + \delta P_{2\mathbf{k_{L}}}^{in} \nonumber
\end{eqnarray}

We can now inject the mean values of the fields $p_{\mathbf{k_{L}}}$, $p_{\mathbf{0}}$ et $p_{2\mathbf{k_{L}}}$ that we
have computed in the preceding section (equations (\ref{intensite seuil}), (\ref{intensite signal}) and (\ref{intensite
compl})).

First we have to choose the phases of the fields (this choice has no influence on the physics of the problem). We set the
phase of the pump field $A_{\mathbf{k_{L}}}^{in}$ to zero. Then  $\overline{p}_{\mathbf{k_{L}}}$ is a positive real
number. The equations (\ref{stat signal}) and (\ref{stat compl}) impose the same relationship between the signal $\varphi
_{\mathbf{0}}$ and idler $\varphi _{2\mathbf{k_{L}}}$ phases:

\begin{equation}
\varphi _{\mathbf{0}}+\varphi _{2\mathbf{k_{L}}}=-\frac{\pi }{2}
\end{equation}

whereas the relative phase $\varphi _{\mathbf{0}}-\varphi _{2\mathbf{k_{L}}}$ is a free parameter. We set
$\overline{p}_{\mathbf{0}}$ to be a real positive number (again, this choice is of no consequence regarding the physics of
the problem). Then $\overline{p}_{2\mathbf{k_{L}}}$ is a pure imaginary number. With these choices of phase, the evolution
equations write

\begin{eqnarray}
\frac{d\delta p_{\mathbf{k_{L}}}}{dt} &=& -\gamma _{k_{L}}\left( \delta p_{\mathbf{k_{L}}} + \left( \sigma -1\right)
\delta p_{\mathbf{k_{L}}}^{\dagger }\right) \nonumber \\&& - \sqrt{2\gamma _{k_{L}}\gamma _{0}\left( \sigma -1\right)
}\delta p_{\mathbf{0}} \\&& -i\sqrt{ 2\gamma
_{k_{L}}\gamma_{2k_{L}}\left( \sigma -1\right) }\delta p_{2\mathbf{k_{L}}} + \delta P_{\mathbf{k_{L}}}^{in} \nonumber \\
\frac{d\delta p_{\mathbf{0}}}{dt} &=& -\gamma _{0}\delta p_{\mathbf{0}} + \sqrt{2\gamma _{k_{L}}\gamma _{0}\left( \sigma
-1\right) }\delta p_{\mathbf{k_{L}}} \nonumber \\&& - i\sqrt{\gamma_{0}\gamma _{2k_{L}}}\delta
p_{2\mathbf{k_{L}}}^{\dagger }
+ \delta P_{\mathbf{0}}^{in} \\
\frac{d\delta p_{2\mathbf{k_{L}}}}{dt} &=& - \gamma _{2k_{L}}\delta p_{2\mathbf{k_{L}}} - i\sqrt{2\gamma _{k_{L}}\gamma
_{2k_{L}}\left( \sigma -1\right) }\delta p_{\mathbf{k_{L}}} \nonumber \\&& - i\sqrt{\gamma _{0}\gamma _{2k_{L}}}\delta
p_{\mathbf{0}}^{\dagger } + \delta P_{2\mathbf{k_{L}}}^{in}
\end{eqnarray}

Thanks to these three equation and their conjugate equations we can calculate the output fluctuations of the pump, signal
and idler fields as a function of the input fluctuations.

\subsection{Amplitude fluctuations}

In this paper we are mostly interested in the amplitude correlations between signal and idler. We will see that in the
simple case where we neglect the renormalization effects it is enough to solve a system of three equations. We define the
real and imaginary parts of the polariton, photon and exciton fields

\begin{eqnarray}
\alpha _{\mathbf{q}} &=&\delta p_{\mathbf{q}}+\delta p_{\mathbf{q}}^{\dagger } \nonumber \\
\beta _{\mathbf{q}} &=&-i\left( \delta p_{\mathbf{q}}-\delta p_{\mathbf{q}}^{\dagger }\right) \\
\alpha _{\mathbf{q}}^{in(out)} &=&\delta P_{\mathbf{q}}^{in(out)}+\delta P_{\mathbf{q}}^{in(out)\dagger } \nonumber \\
\beta _{\mathbf{q}}^{in(out)} &=&-i\left( \delta P_{\mathbf{q}}^{in(out)}-\delta P_{\mathbf{q}}^{in(out)\dagger }\right) \nonumber \\
\alpha _{\mathbf{q}}^{A,in(out)} &=&\delta A_{\mathbf{q}}^{in(out)}+\delta A_{\mathbf{q}}^{in(out)\dagger } \nonumber \\
\beta _{\mathbf{q}}^{A,in(out)} &=&-i\left( \delta A_{\mathbf{q}}^{in(out)}-\delta A_{\mathbf{q}}^{in(out)\dagger }\right) \nonumber \\
\alpha _{\mathbf{q}}^{B,in(out)} &=&\delta B_{\mathbf{q}}^{in(out)}+\delta B_{\mathbf{q}}^{in(out)\dagger } \nonumber \\
\beta _{\mathbf{q}}^{B,in(out)} &=&-i\left( \delta B_{\mathbf{q}}^{in(out)}-\delta B_{\mathbf{q}}^{in(out)\dagger }\right)
\nonumber
\end{eqnarray}

The mean fields $p_{\mathbf{k_{L}}}$ and $p_{\mathbf{0}}$ are real positive numbers, therefore $\alpha$ corresponds to
amplitude fluctuations and $\beta $ to phase fluctuations. The mean field $p_{2\mathbf{k_{L}}}$ is a pure imaginary
number, therefore $-\beta$ corresponds to amplitude fluctuations and $\alpha$ to phase fluctuations. The evolution
equations for the amplitude fluctuations write

\begin{eqnarray}
\frac{d\alpha _{\mathbf{k_{L}}}}{dt} &=& -\gamma_{k_{L}} \sigma \alpha _{\mathbf{k_{L}}} -
\sqrt{2\gamma_{k_{L}}\gamma_{0}\left( \sigma -1\right) }\alpha _{\mathbf{0}} \nonumber \\&&
+\sqrt{2\gamma_{k_{L}}\gamma_{2k_{L}}\left( \sigma-1\right)}\beta _{2\mathbf{k_{L}}}
+ \alpha _{\mathbf{k_{L}}}^{in} \\
\frac{d\alpha _{\mathbf{0}}}{dt} &=& -\gamma_{0} \alpha _{\mathbf{0}}+\sqrt{2\gamma_{k_{L}}\gamma_{0}\left( \sigma
-1\right) }\alpha _{\mathbf{k_{L}}} \nonumber \\&& -\sqrt{\gamma_{0}\gamma_{2k_{L}}}\beta _{2\mathbf{k_{L}}}
+ \alpha _{\mathbf{0}}^{in} \\
\frac{d\beta _{2\mathbf{k_{L}}}}{dt} &=&-\gamma_{2k_{L}} \beta
_{2\mathbf{k_{L}}}-\sqrt{2\gamma_{k_{L}}\gamma_{2k_{L}}\left( \sigma-1\right)}\alpha _{\mathbf{k_{L}}} \nonumber \\&&
-\sqrt{\gamma_{0}\gamma_{2k_{L}}}\alpha _{\mathbf{0}} + \beta _{2\mathbf{k_{L}}}^{in}
\end{eqnarray}

We get a set of three linear differential equations. Taking the Fourier transform we obtain in matrix notation

\begin{widetext}

\begin{equation}
\left(
\begin{array}{ccc}
\gamma_{k_{L}} \sigma -i\Omega & \sqrt{2\gamma_{k_{L}}\gamma_{0}\left(
\sigma -1\right) } & -\sqrt{2\gamma_{k_{L}}\gamma_{2k_{L}}\left( \sigma -1\right) } \\
-\sqrt{2\gamma_{k_{L}}\gamma_{0}\left(\sigma -1\right)} & \gamma_{0} -i\Omega &
\sqrt{\gamma_{0}\gamma_{2k_{L}}}\\
\sqrt{2\gamma_{k_{L}}\gamma_{2k_{L}}\left( \sigma -1\right)} & \sqrt{\gamma_{0}\gamma_{2k_{L}}} & \gamma_{2k_{L}}-i\Omega
\end{array}
\right) \times \left(
\begin{array}{c}
\alpha _{\mathbf{k_{L}}}\left( \Omega \right) \\
\alpha _{\mathbf{0}}\left( \Omega \right) \\
\beta _{2\mathbf{k_{L}}}\left( \Omega \right)
\end{array}
\right) = \left(
\begin{array}{c}
\alpha_{\mathbf{k_{L}}}^{in} \\
\alpha_{\mathbf{0}}^{in} \\
\beta _{2\mathbf{k_{L}}}^{in}
\end{array}
\right) \label{eqmatricielle}
\end{equation}

\end{widetext}

The inversion of the $3 \times 3$ matrix provides the amplitude fluctuations of the fields $p_{\mathbf{k_{L}}}$,
$p_{\mathbf{0}}$ et $p_{2\mathbf{k_{L}}}$ as a function of the input fluctuations. It is easy to deduce the amplitude
fluctuations $\alpha_{\mathbf{q}}^{A,\; out}$ of the output light fields thanks to the input-output relationship for the
cavity mirror $A_{\mathbf{q}}^{out}=\sqrt{2\gamma_{aq}} a_{\mathbf{q}} - A_{\mathbf{q}}^{in}$ and the relationship between
the photon and polariton fields $a_{\mathbf{q}}=-C_{q} p_{\mathbf{q}}$.

\begin{equation}
\alpha_{\mathbf{q}}^{A,\; out} = -C_{q} \sqrt{2\gamma_{aq}} \alpha_{\mathbf{q}}- \alpha_{\mathbf{q}}^{A,\; in}
\end{equation}

\subsection{Input fluctuations}

In this paragraph we study the noise sources in our system. $A_{\mathbf{k_{L}}}^{in}$ is the coherent pump laser field
$A_{\mathbf{0}}^{in}$ and both other input fields $A_{2\mathbf{k_{L}}}^{in}$ are equal to the vacuum field. Therefore, the
amplitude fluctuations of these three fields are equal to the vacuum fluctuations. The treatment of excitonic fluctuation
is more complex. The amplitude noise spectra (normalized to the vacuum noise) of the three excitonic fields
$B_{\mathbf{k_{L}}}^{in}$, $B_{\mathbf{0}}^{in}$ et $B_{2\mathbf{k_{L}}}^{in}$ are given by

\begin{equation}
S_{\alpha_{\mathbf{q}}}^{B,\; in}(\Omega) = 1 + 2 n_{\mathbf{q}} \mbox{ for } \mathbf{q}=\mathbf{0},\mathbf{k_{L}} \;
,2\mathbf{k_{L}}
\end{equation}

where $n_{\mathbf{q}}$ is the mean number of excitations in the reservoir which depends on the temperature and on the pump
intensity. Since the reservoir is populated through phonon-assisted relaxation from the pump mode it is a reasonable
assumption to take the reservoir occupation as proportional to the mean number of excitons in the pump mode:

\begin{equation}
n_{\mathbf{q}} = \beta |b_{\mathbf{q}}|^{2} = \beta X_{q}^{2} |p_{\mathbf{q}}|^{2}
\end{equation}

where $\beta$ is a dimensionless constant which characterizes the efficacy of the relaxation process. This simple model
accounts for the excess noise of the reflected light at low excitation intensity in a satisfactory way \cite{karr}.

\subsection{Noise spectra}

In fluctuation measurements the measured quantity is the noise spectrum. The noise spectrum $S_{O}(\Omega)$ of an operator
$O$ is defined as the Fourier transform of the autocorrelation function $C_{O} \left( t,t' \right)$:

\begin{equation}
S_{O}(\Omega)=\int C_{O}(\tau) e^{i\Omega \tau} d\tau
\end{equation}

with

\begin{equation}
C_{O} \left( t,t' \right) = \left\langle O(t)O(t') \right\rangle - \langle O(t) \rangle \langle O(t') \rangle = \langle
\delta O(t) \delta O(t') \rangle
\end{equation}

The noise spectrum is related to the Fourier transform $\delta O(\Omega)$ of the fluctuations $\delta O(t)$ by the
Wiener-Kinchine theorem:

\begin{equation}
\langle \delta O(\Omega) \delta O(\Omega') \rangle = 2\pi \delta(\Omega +\Omega') S_{O} (\Omega)
\end{equation}

In the same way the correlation spectrum $S_{O \; O'}(\Omega)$ of two operators $O,O'$ is defined as the Fourier transform
of the correlation function:

\begin{equation}
C_{O \; O'} \left( t,t' \right) = \left\langle O(t)O'(t') \right\rangle - \langle O(t) \rangle \langle O'(t') \rangle
\end{equation}

The correlation spectrum is also related to the Fourier components of the fluctuations:

\begin{equation}
\langle \delta O(\Omega) \delta O'(\Omega') \rangle = 2\pi \delta(\Omega +\Omega') S_{O \; O'} (\Omega)
\end{equation}

The relevant quantity is the normalized correlation spectrum

\begin{equation}
C_{O \; O'}(\Omega) = \frac{S_{O \; O'}(\Omega)}{\sqrt{S_{O}(\Omega) S_{O'}(\Omega)}} \label{normal}
\end{equation}

One has always $|C| \leq 1$. A nonzero value of $C_{O \; O'}(\Omega)$ indicates some level of correlation between the two
measurements.

\section{Results}

\subsection{Fluctuations of the intracavity polariton fields}

First, in order to shed some light on the above-mentioned analogy with an OPO, we assume that all three polariton modes
have the same linewidths. This is the case if the cavity and exciton linewidths are equal ($\gamma _{ak}=\gamma _{bk}$)
and do not depend on $k$. We set $\gamma =\gamma _{k_{L}}=\gamma _{0}=\gamma _{2k_{L}}=\gamma _{a}=\gamma _{b}$.

After some straightforward algebra we get the amplitude fluctuations of the polariton fields

\begin{eqnarray}
\alpha_{\mathbf{k_{L}}}\left( \Omega \right) &=& \frac{1}{D\left(\Omega \right)} \left( -\gamma \left( \Omega + 2i\gamma
\right) \alpha_{\mathbf{k_{L}}}^{in} \right. \nonumber \\&& \left. - \gamma \sqrt{2\left( \sigma -1\right) }\left( 2\gamma
-i\Omega \right) \alpha_{\mathbf{0}}^{in} \right. \\&& \left. + \gamma \sqrt{2\left( \sigma -1\right) }\left( 2\gamma
-i\Omega \right) \beta_{2\mathbf{k_{L}}}^{in} \right)
\nonumber \label{polariton1} \\
\alpha _{\mathbf{0}}\left( \Omega \right) &=& \frac{1}{D\left( \Omega \right) } \left( \gamma \sqrt{2\left( \sigma
-1\right) }\left( 2\gamma -i\Omega \right) \alpha_{\mathbf{k_{L}}}^{in} \right. \\&& \left. +(\gamma^{2}\left( 3\sigma
-2\right) -\Omega ^{2}-i\gamma \Omega \left( \sigma +1\right)) \alpha_{\mathbf{0}}^{in} \right. \nonumber \\&& \left.
+\gamma\left( \gamma\left(
\sigma -2\right)+i\Omega \right)\beta_{2\mathbf{k_{L}}}^{in} \right) \nonumber \\
 \beta _{2\mathbf{k_{L}}}\left( \Omega
\right) &=&\frac{1}{D\left( \Omega \right) } \left(  -\gamma \sqrt{2\left( \sigma -1\right) }\left( 2\gamma -i\Omega
\right) \alpha_{\mathbf{k_{L}}}^{in} \right. \nonumber \\&& \left. +\gamma \left( \gamma\left( \sigma - 2\right) +i\Omega
\right) \alpha_{\mathbf{0}}^{in} \right.
\\&& \left. +(\gamma ^{2}\left( 3\sigma -2\right) -\Omega ^{2}-i \gamma \Omega \left( \sigma +1\right))
\beta_{2\mathbf{k_{L}}}^{in} \right) \nonumber \label{polariton3}
\end{eqnarray}

with

\begin{equation}
D\left( \Omega \right) = \gamma \left[ 8\gamma ^{2}\left( \sigma -1\right) -\Omega ^{2}\left( \sigma +2\right) \right]
+i\Omega \left[ \gamma ^{2}\left( 4-6\sigma \right) +\Omega ^{2}\right]
\end{equation}

\subsection{Twin polaritons}

Let us now calculate the fluctuations of the difference of the signal and idler amplitudes. Let $r$ be the normalized
quantity

\begin{equation}
r=\frac{1}{\sqrt{2}}\left(\alpha _{\mathbf{0}}+\beta_{2\mathbf{k_{L}}} \right)
\end{equation}

We find

\begin{eqnarray}
r\left( \Omega \right) &=& \left( 4\gamma^{2} \left( \sigma -1 \right) - \Omega^{2} - i \Omega \gamma \sigma \right)
r^{in} \\ \text{avec }r^{in} &=& \frac{1}{\sqrt{2}} \left( \alpha_{\mathbf{0}}^{in} + \beta_{2\mathbf{k_{L}}}^{in} \right)
\nonumber
\end{eqnarray}

It is important to notice that $r$ does not depend on the pump fluctuations, which cancel out when we make the difference.
This property is at the origin of twin beams generation in OPOs. We get perfect noise suppression for $\Omega=0$ and
$\sigma \rightarrow 1$.

In a degenerate or quasi-degenerate OPO the symmetry between signal and idler is conserved outside the cavity, because the
two fields have the same frequency and are coupled in the same way to the external field through the losses of the cavity
mirrors. In such systems the "twinity" of the signal and idler fields can be shown directly by measuring the fluctuations
of the difference of the output signal and idler field intensities.

In our case the signal and idler polaritons do not have the same photon fraction and are not coupled in the same way to
the external field. Clearly, this should lead to a significant reduction of the correlations between the signal and idler
output light fields.

\subsection{Fluctuations of the output light fields}

Let us first comment on the relevant analysis frequency of the noise. The noise spectra vary typically over a range of the
order of the polariton linewidth. In noise measurements the experimentalists have access to very small analysis
frequencies (generally a few tens of MHz, i.e. a fraction of $\mu$eV) with respect to the polariton linewidths (a few
hundreds of $\mu$eV). Therefore the noise at zero frequency is the relevant quantity.

The general expressions of the amplitude noises of the three modes and of the signal-idler amplitude correlation can be
found in the Appendix.

In expressions (\ref{polariton1}-\ref{polariton3}) we have taken equal linewidths for the pump, signal and idler
polaritons ($\gamma_{k_{L}}=\gamma_{0} =\gamma_{2k_{L}}$). This assumption is not correct in most microcavity samples.
Indeed the energy of the polaritons of wave vector $2\mathbf{k_{L}}$ is close to the energy of the nonradiative excitons ;
diffusion toward these states is enhanced by their large density of states. Moreover, the idler energy is closer to the
electron-hole continuum. As a result, the \textit{excitonic} linewidth of the idler $\gamma_{b2k_{L}}$ is larger than that
of the signal $\gamma_{b0}$ and pump $\gamma_{bk_{L}}$ modes. The assumption that the \textit{cavity} linewidth
$\gamma_{ak}$ does not depend on $k$ is correct provided the three wave vectors of interest are within the stop-band of
the Bragg reflectors. In recent experiments, the idler beam has been found to be about 50 times weaker than the signal
beam (see e.g. Ref.~\cite{baumberg}), which is consistent with a linewidth ratio $\gamma_{2k_{L}}/\gamma_{0} = 5$.

We will first give the results in the ideal case (with equal linewidths and an input noise equal to the standard quantum
noise), and then study the influence of the imbalance between signal and idler and the input excitonic noise.

\subsubsection{Ideal case}

\begin{figure}[p]
\centerline{\includegraphics[clip=,width=16cm]{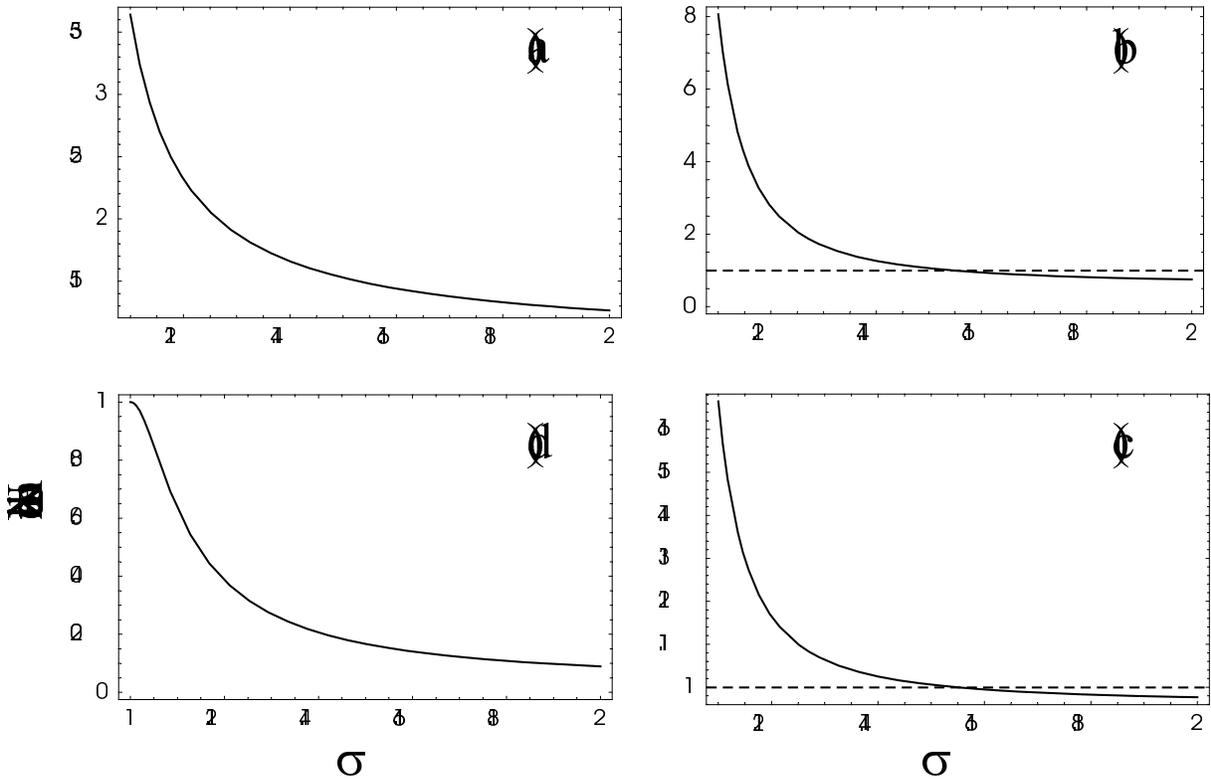}} \caption{(a) through (c) : amplitude noises at zero frequency of
the pump, signal and idler beams respectively. (d) : signal-idler amplitude correlation at zero frequency. The three modes
are assumed to have the same linewidths, and the input noise is set as equal to the standard quantum noise. }
\label{ideal}
\end{figure}

The amplitude noises of the pump, signal and idler beams as well as the signal-idler amplitude correlation are drawn in
Fig.~\ref{ideal} as a function of the pump parameter $\sigma = \sqrt{I_{\mathbf{k_{L}}}^{in} / I_{\mathbf{k_{L}},S}^{in}}$
in the case of equal linewidths (and no input excess noise. Although the curves go up to $\sigma$=5 let us recall that the
model is not correct too far above threshold where we can no longer neglect multiple diffusions.

Let us first observe that the signal and idler noise spectra have exactly the same shape. However the idler noise is drawn
towards the standard quantum level due to its low photon fraction which causes important losses at the output of the
cavity. It is easy to show that the ratio of the noise signals $S-1$ is equal to the ratio of the photon fractions:

\begin{equation}
\frac{S_{\alpha_{\mathbf{0}}}^{A,\; out}(\Omega)-1}{S_{\beta_{2\mathbf{k_{L}}}}^{A,\;
out}(\Omega)-1}=\frac{C_{0}^{2}}{C_{2k_{L}}^{2}}
\end{equation}

The signal and idler amplitude fluctuations diverge close to the threshold (for $\sigma \longrightarrow 1^{+}$). Noise
reduction is obtained above $\sigma =1.55$. It grows with the pump intensity and saturates at a value .... The amplitudes
of the signal and idler beams are very strongly correlated slightly above threshold. The correlation tends to one in the
vicinity of the threshold ($\sigma\rightarrow 1^{+}$) and vanishes rapidly when increasing the pump intensity. All these
results are similar to those obtained in nondegenerate OPOs \cite{opo1}.

\subsubsection{Influence of the signal-idler imbalance}

In this paragraphe we still suppose that there is no input excess noise
($n_{\mathbf{0}}=n_{\mathbf{k_{L}}}=n_{2\mathbf{k_{L}}}=0$). Let us compare the results with different linewidths to those
of the "balanced" case ($\gamma =\gamma _{k_{L}}=\gamma _{0}=\gamma _{2k_{L}}=\gamma _{a}=\gamma _{b}$) in equations
(\ref{pompe2})-(\ref{correl2}). It is easy to show that the excess $S-1$ noises of the pump, signal and idler beams are
respectively multiplied by $\gamma_{a}/\gamma_{k_{L}}$, $\gamma_{a}/\gamma_{0}$ and $\gamma_{a}/\gamma_{2k_{L}}$. The
signal-idler correlation (without normalization) is multiplied by $\gamma_{a}/\sqrt{\gamma_{0} \gamma_{2k_{L}}}$.

\begin{figure}[p]
\centerline{\includegraphics[clip=,width=16cm]{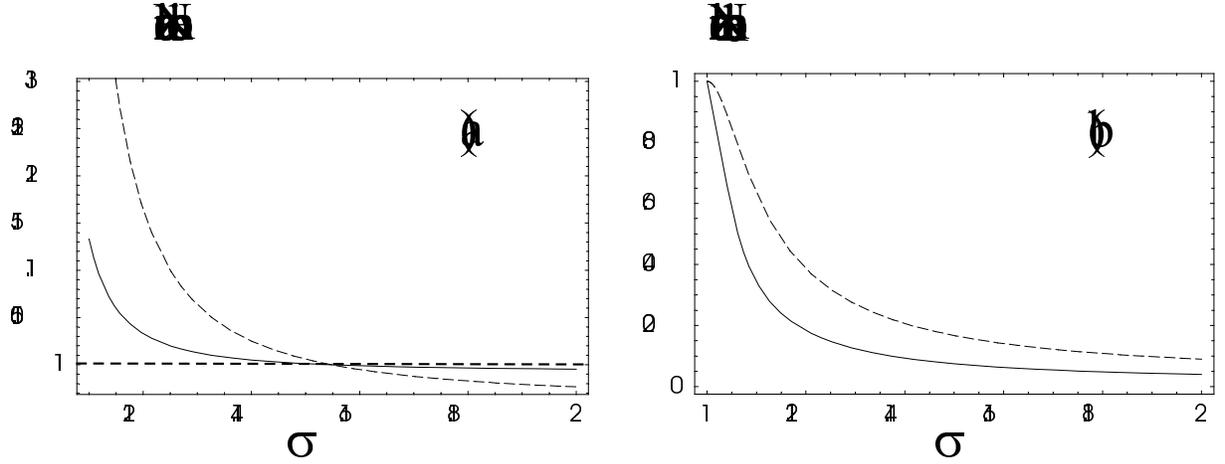}} \caption{(a) Amplitude noise of the idler beam and (b)
signal-idler correlation at zero frequency as a function of pump intensity for $\gamma_{2k_{L}}=5 \gamma_{0}$. On both
plots, the curve in dashed line is the "balanced" case $\gamma_{2k_{L}}=\gamma_{0}$.} \label{desequilibre}
\end{figure}

As an example the case $\gamma_{0}=\gamma_{k_{L}}=\gamma_{2k_{L}}/5=\gamma_{a}$ is shown in Fig.~\ref{desequilibre}. The
amplitude noises of the pump and signal beams have not been represented since they are unchanged. The excess noise and the
noise reduction are strongly reduced on the idler beam due to its larger losses (Fig.~\ref{desequilibre} (a)). The
signal-idler correlation remains strong close to threshold but decreases more rapidly with increasing pump intensity
(Fig.~\ref{desequilibre} (b)).

\subsubsection{Influence of input excess noise}

We have assumed that the biggest source of noise for a given polariton mode is the luminescence of an exciton reservoir
which is populated by the polariton mode itself. The input noise for a given mode is then proportional to the mean exciton
number in this mode. The efficacy of this process is given by the $\beta$ coefficient introduced above ; here we will
assume that $\beta$ has the same value for the three modes. Slightly above the oscillation threshold, the pump mode is
much more populated than the signal and idler population ; then the input noise is much greater for the pump than for the
signal and idler.

\begin{figure}[p]
\centerline{\includegraphics[clip=,width=16cm]{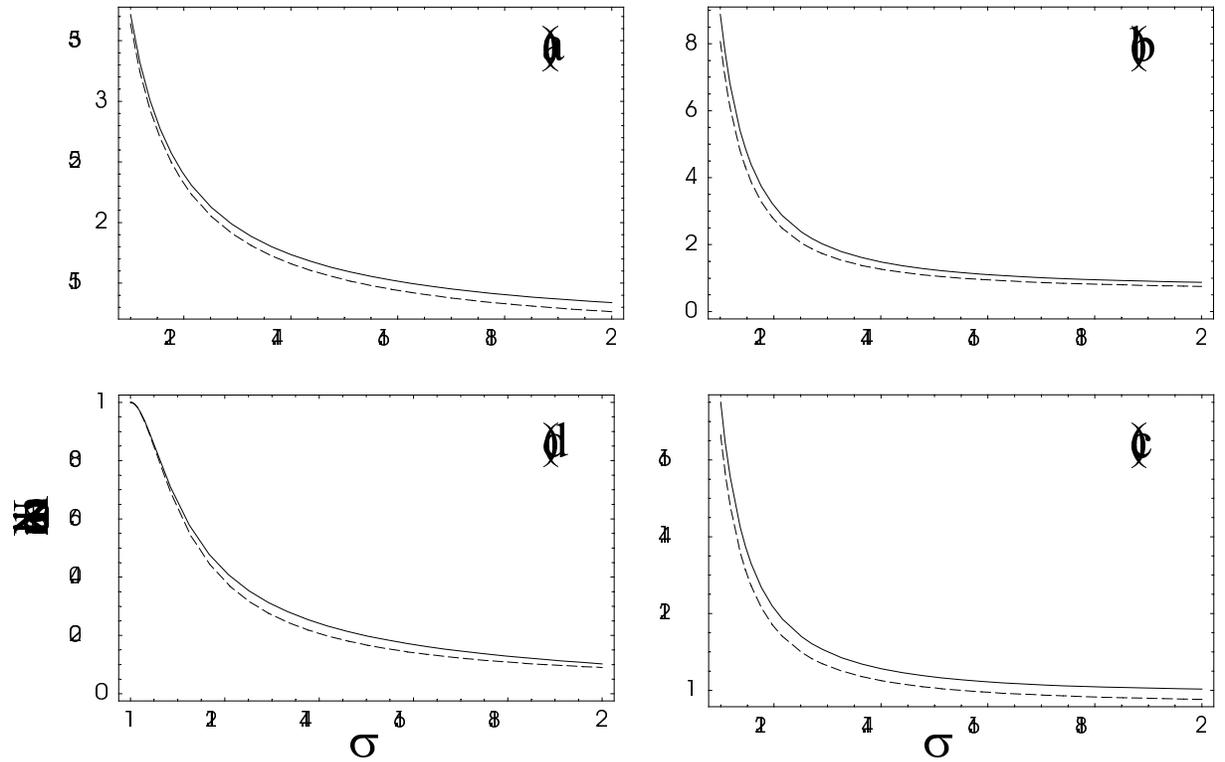}} \caption{Noises at zero frequency in the case of equal
linewidths, with an input excess noise given by $\beta$=$\beta_{c}$/2. The ideal case $\beta$=0 is represented on each
curve as a dashed line. (a) pump beam amplitude noise ; (b) signal beam amplitude noise ;(c) idler beam amplitude noise ;
(d) signal-idler amplitude correlation.} \label{excesbruit}
\end{figure}

Fig.~\ref{excesbruit} shows an example in the "balanced" case for a noise parameter $\beta$=5.10$^{-5}$, evaluated from
noise measurements on the light reflected by a microcavity sample \cite{karr}. The input excess noise cuts down the noise
reduction. Its influence increases with the pump intensity since it is proportional to the mean exciton population.
However the correlation is actually enhanced by the excess noise. It is due to the fact that the pump input noise is
distributed equally between signal and idler and contributes to the correlations.

\subsection{The quantum domain}

Our model predicts strong correlations between the signal and idler light fields. When can we say that these beams are
quantum correlated ? We will use two different criteria, one of "quantum twinity" and one associated with QND measurement.

\subsubsection{Quantum twinity}

In degenerate or quasi degenerate OPOs, the signal and idler output beams have the same mean field values and the same
noise properties. Quantum correlations between them are evidenced by measuring the noise of the difference between signal
and idler intensities and comparing it to the standard quantum level. The idea behind this is to compare the fields under
consideration to a classical production of twin beams, which can be achieved by using a $50\%$ beamsplitter.

In our case, one beam is much more intense than the other one (the ratio of the intensities is of the order of 10 for
equal signal and idler linewidths). What happens if the two light fields $A_{1}$ and $A_{2}$ under consideration have
different mean values and different noises $S_{1}$ and $S_{2}$? To produce classically twin beams of unequal intensities,
one can use an unequal beamsplitter. The field fluctuations at the output of such a beamsplitter write

\begin{eqnarray}
\delta A_{1} &=& t \delta A_{in} + r \delta A_{v} \\
\delta A_{2} &=& r \delta A_{in} - t \delta A_{v}
\end{eqnarray}

with $t \neq r$, where $A_{in}$ is the input field and $\delta A_{v}$ the vacuum fluctuations entering through the other
port of the beamsplitter. Now the difference $\delta A_{-} = \delta A_{1} - \delta A_{2}$ is not helpful, since it does
not give a quantity which is independent of $\delta A_{in}$. However, the correlation between the two beams is independent
of $\delta A_{in}$:

\begin{equation}
\langle \delta A_{1} \delta A_{2} \rangle^{2}_{classical twins} = \left( \langle \delta A_{1}^{2} \rangle - 1 \right)
\left( \langle \delta A_{2}^{2} \rangle - 1 \right)
\end{equation}

which can also be written as:

\begin{equation}
(C_{classical twins})^{2} = \left( 1 - \frac{1}{S_{1}} \right) \left( 1 - \frac{1}{S_{2}} \right)
\end{equation}

We can evaluate the "twinity" of the beams by using the quantity

\begin{equation}
G = \frac {1-C} {1-\sqrt{\left(1-\frac{1}{S_{1}}\right) \left(1-\frac{1}{S_{2}}\right)}} \label{jumeaux}
\end{equation}

which is a generalization of the usual squeezing factor on the intensity difference. $G$ smaller than 1 means that one has
been able to produce two fields which are more identical than the copies from a beamsplitter. Moreover, it is possible to
show that this criterium does not depend on the way by which the two classical twins are produced \cite{claude}.

Experimentally, one can measure separately $C$, $S_{1}$ and $S_{2}$ and compute $G$ from (\ref{jumeaux}). One can also
amplify in a different way the two photocurrents in order to measure the quantity $\delta A_{a} = a \delta A_{1} - \delta
A_{2}/a$. When $a^{2} = \sqrt{S_{2}/S_{1}}$ then:

\begin{equation}
G = \frac{\langle \delta A_{a}^{2} \rangle}{2} \frac{1}{\sqrt{S_{1} S_{2}}-\sqrt{(S_{1}-1)( S_{2}-1)}} \label{jumeaux2}
\end{equation}

G is proportional to the photocurrent fluctuations when the gains are adjusted so that the noise levels are identical in
the two channels. The denominator in (\ref{jumeaux2}) can be evaluated from the excess noises of each field.

\subsubsection{QND correlation}

A further level of correlation is achieved when the information extracted from the measurement of one field provides a QND
measurement of the other, so that it is possible, using the information on one field, to correct the other from a part of
its quantum fluctuations and transform it into a squeezed state. This criterium is widely used in the field of QND
measurement \cite{poizat}. It can be expressed in terms of the conditional variance

\begin{equation}
V_{1|2} = S_{1} (1-C^{2})
\end{equation}

Note that when the two beams have different noises ($S_{1} \neq S_{2}$) one has two conditional variances and therefore
two possible criteria. This shows that the QND criterium evaluates the correlation from the point of view of one beam, and
is not an evaluation of the quantum correlation between the two fields. One possibility is to state that the two fields
are QND correlated if one has $V_{1|2}<1$ \textit{and} $V_{2|1}<1$. This criterium is stronger than the previous one
\cite{claude}.

\subsubsection{Discussion}

\begin{figure}[p]
\centerline{\includegraphics[clip=,width=8cm]{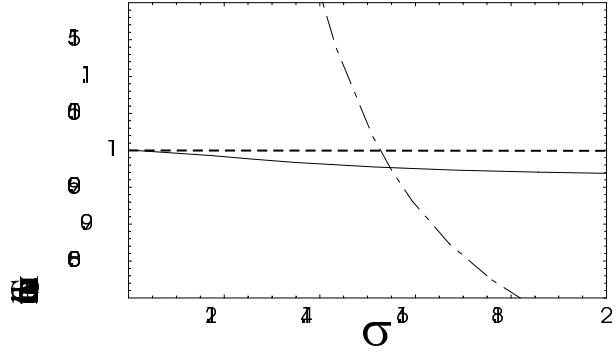}} \caption{(Dash-dotted line : conditional variance of the signal
intensity fluctuations, knowing those of the idler ; solid line : conditional variance of the idler intensity
fluctuations, knowing those of the signal. The dashed line is the standard quantum level} \label{quantique}
\end{figure}

We first investigate the "QND criterium". The conditional variances are shown in Fig.~\ref{quantique} in the case of equal
linewidths and zero input excess noise. From the point of view of the idler beam, the conditional variance is always lower
than 1, if only by a few percent. From the point of view of the signal beam, the quantum domain is very small: it begins
at $\sigma$ = 1.53, very close to the point where it begins to be squeezed. It is only between $\sigma$=1.53 and
$\sigma$=1.55 that we get "QND correlations" between beams that individually have excess noise. For $\sigma >$ 1.55 the
QND correlation criterium is satisfied, although the correlation is quite small, because both beams are squeezed. In
conclusion, no significant "QND correlations" can be observed on the signal and idler output beams.

\begin{figure}[p]
\centerline{\includegraphics[clip=,width=8cm]{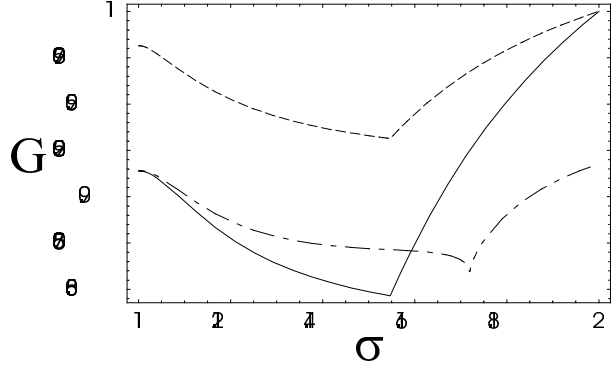}} \caption{Value of the "gemellity" $G$ as a function of the pump
parameter, in three different cases. (a) solid line : ideal case where all linewidths are equal and there is no excess
noise. (b) dashed line : different linewidths for the signal and idler modes $\gamma_{2k_{L}}$= 5 $\gamma_{0}$, and no
excess noise. (c) dashed-dotted line : all linewidths equal, and some excess noise given by $\beta$=$\beta_{c}$/2.}
\label{realiste}
\end{figure}

We now investigate the behavior of the quantity $G$ evaluating the "twinity" of the signal and idler beams. It is drawn in
Fig.~\ref{realiste} as a function of the pump parameter in various cases. In the case of equal linewidths and zero input
excess noise, $G$ goes down to 0.85 which indicates the "quantum twin" character of the two beams. If we take the
nonradiative losses of the idler polaritons into account (we set again $\gamma_{2k_{L}} = 5 \gamma_{0}$) $G$ only goes
under 1 by a 7 percent. However the input excess noise (corresponding to the resonant luminescence of the three polariton
modes) has little effect on the quantum correlations. As explained above this comes from the fact that the pump input
noise (which is the strongest slightly above threshold, when the pump polariton population is much larger than the signal
and idler populations) is equally distributed between the signal and idler modes and helps building up correlations.

In conclusion, in present-day microcavity samples the "quantum twinity" criterium is overcome by only a few percent. This
is due to the fact that only the polariton fields are perfectly correlated, and we can only observe their photonic parts.
A simple image is the following: we observe the polariton system through a beamsplitter which amplitude transmission
coefficient is equal to the Hopfield coefficient $C_{0}$, which leads to losses that destroy the quantum effects. The
correlations are further reduced by the imbalance between signal and idler. The photonic part of the idler is very small
(of the order of 0.05) which corresponds to large losses.

\section{Conclusion}

We have presented a novel quantum model allowing to calculate the quantum fluctuations of the beams produced by a
semiconductor microcavity in the regime of parametric oscillation. It extends the model developed by C. Ciuti et al. above
threshold and includes the noise coming from the exciton part of the polaritons.

We show that some quantum correlation exists between the signal and idler beams in the vicinity of threshold. Taking the
parameters of microcavity samples which have been shown to work in the parametric oscillation regime, it can be seen that
the correlation overcomes the quantum limit by a few percent. The measurement of these correlations would be of great
interest, since quantum correlations between the output beams, however small, are an indication of much bigger
correlations between the intracavity polariton fields. For example, in the ideal case at threshold (see
Fig.~\ref{gemellity}), if we measure a gemellity G=0.91 this corresponds to perfect correlations inside the cavity.

In order to observe better quantum correlations between the output beams, it is very important that the signal and idler
linewidths should be made as equal as possible. A simple solution would be to use a low finesse cavity. Then the
nonradiative losses would be less important with respect to the radiative losses and the ratio of the signal and idler
linewidths would be smaller. A compromise has to be found because the oscillation threshold would also be higher.

We acknowledge fruitful discussions with C. Fabre, C. Ciuti, P. Schwendimann and A. Quattropani.

\section{APPENDIX : Noise and signal-idler correlation}

In this paragraph we give the general expressions for the amplitude noises of the signal, pump and idler output light
fields at zero frequency (denoted by $S_{\alpha_{\mathbf{0}}}^{A,\; out}$, $S_{\alpha_{\mathbf{k_{L}}}}^{A,\; out}$ and
$S_{\beta_{\mathbf{2 k_{L}}}}^{A,\; out}$ respectively), and the signal-idler amplitude correlation at zero frequency
(denoted by $S_{\alpha_{\mathbf{0}} \; \beta_{2 \mathbf{k_{L}}}}^{A,\; out}$).

\begin{eqnarray}
&& S_{\alpha_{\mathbf{k_{L}}}}^{A,\; out} = 1 + C_{k_{L}}^{2} \frac{\gamma_{a}}{\gamma_{k_{L}}} \frac{1}{\sigma -1} \times \nonumber \\
&& \left[ 1 + \; \frac{ X_{0}^{2} n_{\mathbf{0}} \gamma_{b0} \gamma_{2k_{L}} + X_{2k_{L}}^{2} n_{2\mathbf{k_{L}}}
\gamma_{b2k_{L}} \gamma_{0}} {\gamma_{0} \gamma_{2k_{L}}} \right] \label{pompe2} \\
&& S_{\alpha_{\mathbf{0}}}^{A,\; out} = 1 + C_{0}^{2}
\frac{\gamma_{a}}{\gamma_{0}} \; \frac{1} {8 \left( \sigma - 1 \right)^{2} } \times \nonumber \\
&& \left[ -7\sigma^{2} +16\sigma -8 + \; \frac{1} {\gamma_{k_{L}} \gamma_{0} \gamma_{2k_{L}}} \times \right. \nonumber \\
&& \left( 8\left( \sigma -1 \right) X_{k_{L}}^{2} n_{\mathbf{k_{L}}} \gamma_{bk_{L}} \gamma_{0} \gamma_{2k_{L}} + \left(
3\sigma - 2 \right)^2 X_{0}^{2}
n_{\mathbf{0}} \gamma_{b0} \gamma_{k_{L}} \gamma_{2k_{L}} \right. \nonumber \\
&& \left. \left. + \left( \sigma - 2 \right)^{2} X_{2k_{L}}^{2} n_{2\mathbf{k_{L}}} \gamma_{b2k_{L}} \gamma_{k_{L}} \gamma_{0} \right) \right] \\
&& S_{\beta_{2\mathbf{k_{L}}}}^{A,\; out} = 1 + C_{2k_{L}}^{2}
\frac{\gamma_{a}}{\gamma_{2k_{L}}} \; \frac{1} {8 \left( \sigma - 1 \right)^{2} } \times \nonumber \\
&& \left[ -7\sigma^{2} +16\sigma -8 + \; \frac{1} {\gamma_{k_{L}} \gamma_{0} \gamma_{2k_{L}}} \times \right. \nonumber \\
&& \left( 8\left( \sigma -1 \right) X_{k_{L}}^{2} n_{\mathbf{k_{L}}} \gamma_{bk_{L}} \gamma_{0} \gamma_{2k_{L}} + \left(
\sigma - 2 \right)^2 X_{0}^{2}
n_{\mathbf{0}} \gamma_{b0} \gamma_{k_{L}} \gamma_{2k_{L}} \right. \nonumber \\
&& \left. \left. + \left( 3\sigma - 2 \right)^{2} X_{2k_{L}}^{2} n_{2\mathbf{k_{L}}} \gamma_{b2k_{L}} \gamma_{k_{L}} \gamma_{0} \right) \right] \\
&& S_{\alpha_{\mathbf{0}} \; -\beta_{2\mathbf{k_{L}}}}^{A,\; out}= C_{0} C_{2k_{L}} \frac{\gamma_{a}}{\sqrt{\gamma_{0}
\gamma_{2k_{L}}}} \frac{1}{8\left(
\sigma -1 \right)^2} \times  \left[ \; \sigma^{2} - \right. \nonumber \\
&& \left. \frac{1}{\gamma_{k_{L}} \gamma_{0} \gamma_{2k_{L}}} \times  \left[ \left(
\sigma - 2 \right) \left( 3\sigma - 2 \right) \left( X_{0}^{2} n_{\mathbf{0}} \gamma_{b0} \gamma_{k_{L}} \gamma_{2k_{L}} \right. \right. \right. \\
&& \left. \left. \left. + X_{2k_{L}}^{2} n_{2\mathbf{k_{L}}} \gamma_{b2k_{L}} \gamma_{k_{L}} \gamma_{0} \right) - 8 \left(
\sigma -1 \right) X_{k_{L}}^{2} n_{\mathbf{k_{L}}} \gamma_{bk_{L}} \gamma_{0} \gamma_{2k_{L}} \right] \right] \nonumber
\label{correl2} \end{eqnarray}

where $n_{\mathbf{0}}$, $n_{\mathbf{k_{L}}}$ and $n_{\mathbf{2k_{L}}}$ are the input excitonic noises. From these
expressions, it is easy to calculate the normalized signal-idler correlation at zero frequency $C_{\alpha_{\mathbf{0}} \;
\beta_{2 \mathbf{k_{L}}}}^{A,\; out}(\Omega)$, using definition (\ref{normal}).

\end{document}